# Detection of the Temperature Dependence of the White Dwarf Mass–Radius Relation with Gravitational Redshifts

Nicole R. Crumpler[1,8], Vedant Chandra[2], Nadia L. Zakamska[1], Gautham Adamane Pallathadka[1], Stefan Arseneau[1,4],
Nicola Gentile Fusillo[3], J. J. Hermes[4], Carles Badenes[5,6], Priyanka Chakraborty[2], Boris T. Gänsicke[7], and
Stephen P. Schmidt[1,8]
[1] William H. Miller III Department of Physics and Astronomy, Johns Hopkins University, 3400 N Charles Street, Baltimore, MD 21218, USA; ncrumpl2@jh.edu
[2] Center for Astrophysics | Harvard & Smithsonian, 60 Garden Street, Cambridge, MA 02138, USA
[3] Department of Physics, Universita' degli Studi di Trieste, Via A. Valerio 2, 34127, Trieste, Italy
[4] Department of Astronomy & Institute for Astrophysical Research, Boston University, 725 Commonwealth Avenue, Boston, MA 02215, USA
[5] Department of Physics and Astronomy, University of Pittsburgh, 3941 O'Hara Street, Pittsburgh, PA 15260, USA
[6] Pittsburgh Particle Physics, Astrophysics, and Cosmology Center (PITT PACC), University of Pittsburgh, Pittsburgh, PA 15260, USA
[7] Department of Physics, University of Warwick, Coventry CV4 7AL, UK
Received 2024 August 31; revised 2024 October 15; accepted 2024 October 21; published 2024 December 18

## Abstract

Models predict that the well-studied mass–radius relation of white dwarf stars depends on the temperature of the star, with hotter white dwarfs having larger masses at a given radius than cooler stars. In this paper, we use a catalog of 26,041 DA white dwarfs observed in Sloan Digital Sky Survey Data Releases 1–19. We measure the radial velocity, effective temperature, surface gravity, and radius for each object. By binning this catalog in radius or surface gravity, we average out the random motion component of the radial velocities for nearby white dwarfs to isolate the gravitational redshifts for these objects and use them to directly measure the mass–radius relation. For gravitational redshifts measured from binning in either radius or surface gravity, we find strong evidence for a temperature-dependent mass–radius relation, with warmer white dwarfs consistently having greater gravitational redshifts than cool objects at a fixed radius or surface gravity. For warm white dwarfs, we find that their mean radius is larger and mean surface gravity is smaller than those of cool white dwarfs at $5.2\sigma$ and $6.0\sigma$ significance, respectively. Selecting white dwarfs with similar radii or surface gravities, the significance of the difference in mean gravitational redshifts between the warm and cool samples is $>6.1\sigma$ and $>3.6\sigma$ for measurements binned in radius and surface gravity, respectively, in the direction predicted by theory. This is an improvement over previous implicit detections, and our technique can be expanded to precisely test the white dwarf mass–radius relation with future surveys.

*Unified Astronomy Thesaurus concepts:* DA stars (348); White dwarf stars (1799); Fundamental parameters of stars (555); General relativity (641)

## 1. Introduction

Stars that are not massive enough to turn into neutron stars or black holes at the end of their stellar evolution expel their outer layers, leaving behind their cores as compact remnants known as white dwarfs (WDs). All stars that have initial masses ranging from $\sim 0.07$–$8\,M_\odot$, which is around 97% of all stars (G. Fontaine et al. 2001), end their lives as WDs. WDs have been observed with masses ranging from $\sim 0.2$–$1.3\,M_\odot$ and radii ranging from 0.008–$0.2\,R_\odot$, which result in extremely high densities, on the order of $10^6\,\mathrm{g\,cm^{-3}}$ (D. Saumon et al. 2022). At these densities, the electrons in WDs are so tightly compacted that they exhibit quantum mechanical effects on large scales. Thus, WDs are primarily supported by electron degeneracy pressure.

The structure of a star supported entirely by electron degeneracy pressure has been well studied (C. J. Hansen et al. 2004). For low-mass WDs in the zero-temperature approximation, this structure results in the well-known mass–radius relation

$$\frac{M}{M_\odot} = 2.08 \times 10^{-6} \left(\frac{2}{\mu_e}\right)^5 \left(\frac{R}{R_\odot}\right)^{-3}, \quad (1)$$

where $\mu_e$ is the mean molecular weight per free electron. This relation has the property that more-massive WDs have smaller radii, and are thus significantly more compact than less-massive WDs.

For high-mass WDs, relativistic effects modify Equation (1) to $M \propto R^0$, a constant (C. J. Hansen et al. 2004). As a result, WDs have a maximal mass, which is, famously, the Chandrasekhar mass ($M_C$; S. Chandrasekhar & E. A. Milne 1933). Modern studies of stellar evolution have found that this maximal mass, for single-star evolution, is (K. Takahashi et al. 2013; C. L. Doherty et al. 2015)

$$\frac{M_C}{M_\odot} = 1.37. \quad (2)$$

This upper limit on the masses of WDs has important consequences in fields such as supernova cosmology (A. G. Riess et al. 1998; A. G. Riess 2020).

The WD mass–radius relation has been observationally tested in a variety of studies by combining spectroscopic and photometric observations of a type of WD known as a DA WD

---

[8] NSF Graduate Research Fellow.

(G. Vauclair et al. 1997; A. Bédard et al. 2017; P. E. Tremblay et al. 2017; C. Genest-Beaulieu & P. Bergeron 2019; V. Chandra et al. 2020b). DA WDs are WDs with hydrogen-dominated atmospheres, showing Balmer series hydrogen absorption lines in their spectra. This is the most common and well-understood class of WDs, making up ∼80% of all observed WDs in magnitude-limited samples (S. O. Kepler et al. 2019), and thus is best-suited for statistical studies of WDs requiring precise measurements of the star's physical parameters. Spectroscopic observations of DA WDs can be used to measure the star's apparent radial velocity ($v_{\rm app}$), effective temperature ($T_{\rm eff}$), and surface gravity (log $g$) through comparisons to theoretical spectra such as those produced by P. E. Tremblay et al. (2013) or D. Koester (2010). On the other hand, photometric observations of WDs can be combined with parallax measurements to measure the star's radius ($R$) and effective temperature (P. Bergeron et al. 2019). The star's radius and surface gravity constrain the object's mass and its agreement with the WD mass–radius relation.

Additionally, in a method pioneered by R. E. Falcon et al. (2010) and extended by C. Badenes & D. Maoz (2012) and V. Chandra et al. (2020b), averaging out the random motions of sufficiently nearby WDs can be used to isolate the gravitational redshift of WDs from their radial velocity measurements. Gravitational redshift is a general relativistic effect that describes how light leaving the surface of a compact object is redshifted depending on the object's mass and radius (A. Einstein 1916). Thus, binning in radius or surface gravity and averaging the radial velocities of each bin can average out the random motions to isolate the gravitational redshifts of each bin. This method can be used to constrain the mass–radius relation in a way that is robust against the underlying assumptions of model spectra.

Although to leading order the WD mass–radius relation is well described by pure electron degeneracy, more detailed models predict that the relation does depend somewhat on the temperature of the star, with hotter stars having larger radii at a given mass. Figures 7–10 of D. Saumon et al. (2022) illustrate the change of various important physical parameters as a 0.6 $M_\odot$ hydrogen-atmosphere WD cools from an effective temperature of 30,000 K–3000 K. During this process, the temperature-density profile of the carbon-oxygen core and helium layer change very little, but the profile of the hydrogen atmosphere varies greatly. As the WD cools, the temperature and density of this outer hydrogen layer decrease by orders of magnitude, and the location of the photosphere, which can be thought of as the stellar surface, occurs at greater densities and smaller radii. Similarly, the density profile as a function of enclosed mass illustrates that nearly all of the WD mass is enclosed at higher densities, or smaller radii, for cooler WDs than for warmer WDs.

Although the WD mass–radius relation has been tested observationally, detecting the temperature dependence of this relation has only been accomplished implicitly, due to the degeneracy between hydrogen layer thickness and temperature. Most known WDs have surface temperatures ranging from 8000–16,000 K, for which the core temperatures vary from ∼5 × 10$^6$ to 2 × 10$^7$ K (G. Fontaine et al. 2001). For a 0.6 $M_\odot$ WD, a surface temperature increase from 8000 to 16,000 K corresponds to a 5% radius increase from 0.0127 $R_\odot$ to 0.0132 $R_\odot$ (G. Fontaine et al. 2001). This radius difference is smaller than, but comparable to, the typical uncertainty of WD radius measurements, on the order of 10% for WDs with Gaia $G$-band magnitudes of ∼18. Many studies have implicitly detected the temperature dependence of the WD mass–radius relation by plotting the observed spectroscopic surface gravities of WDs as a function of the temperature, as was employed in A. Gianninas et al. (2010) and A. Bédard et al. (2020). These plots reveal that the surface gravity distribution of hotter WDs is shifted toward lower surface gravities than the corresponding distribution for cooler stars, in line with expectations from theory. An additional method is to use high-precision mass, radius, and temperature measurements of individual WDs to characterize the combined effect of temperature and hydrogen envelope thickness on the WD mass–radius relation. This has been accomplished with WDs in visual binaries (H. E. Bond et al. 2015, 2017a, 2017b), in eclipsing binaries (S. G. Parsons et al. 2017), and in wide binaries (S. Arseneau et al. 2024) as well as with isolated WDs with astrometric microlensing data (K. C. Sahu et al. 2017; P. McGill et al. 2023). Improved statistics from a larger sample of DA WDs capable of measuring the full empirical mass–radius relation can confirm the temperature dependence of the WD mass–radius relation for a wide range of WD masses and isolate the effect from that of the thickness of the hydrogen envelope.

WDs are abundant in our Galaxy, but are relatively small and typically less luminous than other types of stars. The European Space Agency's Gaia Mission and the Sloan Digital Sky Survey (SDSS) have revolutionized the photometry (N. P. Gentile Fusillo et al. 2021) and spectroscopy (S. O. Kepler et al. 2019) of WDs, respectively, enabling precision measurements of tens of thousands of these stars. The latest generations of Gaia and SDSS, the third data release of Gaia (Gaia DR3; Gaia Collaboration et al. 2023), and the fifth generation of SDSS (SDSS-V; J. Kollmeier et al. 2019), respectively, have continued to grow the number of observed WDs while simultaneously improving the quality of available data. These developments have enabled us to improve upon previous implicit detections to directly detect the temperature dependence of the WD mass–radius relation.

In this paper, we use a catalog (Crumpler et al. 2024, in preparation) of measured physical parameters for all DA WDs observed in the 19th Data Release of the ongoing SDSS-V survey and identified in previous generations of SDSS to directly measure the temperature dependence of the WD mass–radius relation. In Section 2, we briefly describe the process of creating our SDSS-V and previous SDSS data release DA WD catalogs. In Section 3, we use these measurements to empirically measure the WD mass–radius relation. First, we perform quality cuts on our sample in Section 3.1. We then describe our averaging procedure to isolate gravitational redshifts in Section 3.2. We correct the results of this averaging for bias effects in Section 3.3. We discuss the impacts of binary system and thin hydrogen layer WD contamination in our catalogs in Sections 3.4 and 3.5, respectively. Finally, we show the agreement between our measurements and theoretical models of the WD mass–radius relation in Section 3.6. In Section 4, we use our catalog to directly detect the temperature dependence of the WD mass–radius relation. We conclude in Section 5. All spectra are on the vacuum wavelength scale. Surface gravities are measured on the log $g$ scale in dex where $g$, the surface gravity, is in CGS units.

## 2. Catalog Construction

In this Section, we briefly describe the selection criteria and measurement processes used in constructing our two catalogs. A companion paper (N. R. Crumpler et al. 2024, in preparation), details the construction and validation of this catalog in depth, and we summarize the results of that paper here. The first catalog consists of 8545 unique DA WDs identified in SDSS Data Release 19, part of the ongoing SDSS-V survey. The second is made up of 19,257 unique DA WDs identified across previous SDSS data releases. These catalogs will be released to the public as part of SDSS Data Release 19 as an official SDSS Value Added Catalog (VAC). All code used to create these catalogs will be made publicly available along with the VAC companion paper (N. R. Crumpler et al. 2024, in preparation).

The SDSS-V catalog is selected from all SDSS-V Data Release 19 objects identified as DA WDs by the spectral classification algorithm included in the SDSS-V pipeline known as SnowWhite through 2023 November. The SDSS-V survey (J. Kollmeier et al. 2019) began operations in 2020 November. Initially, the survey operated only from the 2.5 m telescope located at the Apache Point Observatory in sunspot, New Mexico (J. E. Gunn et al. 2006), but then expanded to the 2.5 m telescope at Las Campanas Observatory in the Atacama Desert of Chile (I. S. Bowen & A. H. J. Vaughan 1973) in 2022. WDs observed in SDSS-V were targeted through the Milky Way Mapper program, which focuses on observing millions of stars in the Galaxy with multi-epoch spectroscopy (J. A. Johnson et al. 2024, in preparation). All SDSS-V WD spectra used in this paper were obtained with the Baryon Oscillation Spectroscopic Survey spectrograph (BOSS; S. A. Smee et al. 2013) using the reduction pipeline v6_1_3. BOSS is a low-resolution spectrograph ($R \sim 1800$) covering wavelengths in the range 3650–9500 Å.

Observations in the previous SDSS catalog were compiled by N. P. Gentile Fusillo et al. (2021), covering WDs observed in SDSS through Data Release 16.[9] N. P. Gentile Fusillo et al. (2021) visually inspected all spectra, providing spectral classifications of each observation, and we select only those objects classified as DA WDs. N. P. Gentile Fusillo et al. (2021) measured the physical properties of these objects using Gaia photometry. We cross-match these observations with previously published SDSS WD catalogs to obtain reference physical properties measured from SDSS spectra, which we use to characterize the systematic uncertainties in our spectral fitting procedures. These catalogs include the S. O. Kepler et al. (2019) Data Release 14, S. O. Kepler et al. (2016) Data Release 12, S. O. Kepler et al. (2015) Data Release 10, S. J. Kleinman et al. (2013) Data Release 7, D. J. Eisenstein et al. (2006) Data Release 4, and S. J. Kleinman et al. (2004) Data Release 1 papers. Data releases from previous generations of SDSS only used the 2.5 m telescope at Apache Point Observatory (J. E. Gunn et al. 2006), and Data Releases 1 through 8 used the original SDSS spectrograph. This low-resolution spectrograph ($R \sim 1800$) covered a wavelength range of 3800–9000 Å (D. G. York et al. 2000).

We cross-reference these SDSS observations of DA WDs with the Gaia DR3 catalog to obtain the Gaia photometry and proper-motion measurements for each WD and the C. A. L. Bailer-Jones et al. (2021) distances of each object.

We also cross-reference all selected SDSS objects with the SDSS Data Release 17 (Abdurro'uf et al. 2022) photometric catalog to obtain the SDSS $ugriz$ photometry. With the SDSS spectra, SDSS and Gaia photometry, and distances of each WD in hand, we measure the radial velocities, temperatures, surface gravities, and radii of all WDs in the catalog.

The apparent radial velocity of a DA WD is given by the observed shift in the centroids of the Balmer absorption lines from the measured centroids of these lines in the vacuum. We fit the H$\alpha$, H$\beta$, H$\gamma$, and H$\delta$ hydrogen Balmer lines to P. E. Tremblay et al. (2013) model spectra using a $\chi^2$-minimization routine built into the open-source code Compact Object Radial Velocities (corv[10]). The P. E. Tremblay et al. (2013) grid[11] consists of three-dimensional pure-hydrogen non-local thermodynamic equilibrium (non-LTE) synthetic spectra appended with their one-dimensional LTE spectra (P.-E. Tremblay & P. Bergeron 2009) at low effective temperatures ($<40,000$ K) and with their one-dimensional non-LTE spectra (P. E. Tremblay et al. 2011) at high effective temperatures ($>40,000$ K). Within both the SDSS-V and previous SDSS sets of observations, we coadd all observations corresponding to each unique WD designated by a unique Gaia DR3 source ID. To coadd the observations, we use a routine that resamples each spectrum so all are on a consistent wavelength grid, and then combines the resulting spectra using a weighted mean. We then fit the radial velocities of each individual SDSS spectrum and each coadded spectrum, and record the results. We validate this fitting method by comparing the measured corv apparent radial velocities to published SDSS WD catalog radial velocities, and find agreement to better than 9 km s$^{-1}$ for spectra with signal-to-noise ratio (SNR) $\geqslant 20$. The absolute wavelength calibration of SDSS-V is $\sim 7$ km s$^{-1}$ (S. Arseneau et al. 2024), meaning this agreement is excellent. For spectra with SNR $< 10$, the uncertainties on the measured radial velocities increase dramatically with the mean uncertainty on radial velocities measured from spectra with $5 <$ SNR $< 10$ being $\sim 40$ km s$^{-1}$.

The shapes of the Balmer series lines in DA WD spectra depend on the effective temperature and surface gravity of the star (P. E. Tremblay et al. 2013). We fit the first six Balmer series lines (H$\alpha$, H$\beta$, H$\gamma$, H$\delta$, H$\epsilon$, H$\zeta$) using a parametric random forest routine built into the publicly available code wdtools[12] (V. Chandra et al. 2020a). This routine is trained on P. E. Tremblay et al. (2019) data, which contain corrections for three-dimensional effects, to build a regression between the structure of the absorption lines and the temperature and surface gravity of DA WDs. We fit the spectroscopic surface gravities and temperatures of each individual SDSS spectrum and each coadded spectrum. We validate this fitting method by comparing the measured wdtools spectroscopic parameters to published SDSS WD catalog values. For surface gravities, we find agreement to better than 0.095 dex and 0.062 dex for spectra with SNR $\geqslant 20$ and $\geqslant 50$, respectively. For effective temperatures, we find agreement to better than 2.4% and 2.3% for spectra with SNR $\geqslant 20$ and $\geqslant 50$, respectively. For context, effective temperatures are generally calculated to within 50–1000 K ($\sim 5$%) and surface gravities to within 0.1 dex (V. Chandra et al. 2020a), meaning our fitting techniques are in agreement with the other catalogs and approach the limiting precision of our ability to determine WD spectral parameters with SDSS BOSS spectra and available models. For spectra with SNR < 10, the uncertainties on the measured spectral parameters increase dramatically with the mean uncertainty on surface gravities and effective temperatures measured from spectra with $5 < \text{SNR} < 10$ being $\sim 0.25$ dex and 10%, respectively.

---

[9] https://cdsarc.cds.unistra.fr/viz-bin/cat/J/MNRAS/508/3877#/browse
[10] https://github.com/vedantchandra/corv
[11] https://warwick.ac.uk/fac/sci/physics/research/astro/people/tremblay/modelgrids/
[12] https://wdtools.readthedocs.io/en/latest/

The measured photometry of a WD depends on the star's distance, radius, effective temperature, and, to a lesser extent, the star's surface gravity. We de-redden the observed SDSS and Gaia magnitudes using the three-dimensional dust map of G. Edenhofer et al. (2024) from the publicly available `dustmaps`[13] Python package (G. M. Green 2018) and the extinction curve from E. L. Fitzpatrick (1999) using the `extinction`[14] Python package. We then use $\chi^2$ minimization to fit the observed photometry to model SDSS $urz$ and Gaia $G_{BP}$ and $G_{RP}$ photometry created by convolving P. E. Tremblay et al. (2013) model spectra through photometric filter response curves from the `pyphot`[15] Python package. For the SDSS photometric fit, we first fit the photometric parameters, leaving $\log g$ free to vary. If the fit fails or hits the edge of the surface gravity grid, we re-fit the stellar parameters fixing the surface gravity of the WD to $\log g = 8$ dex, the peak value in the WD mass distribution. In this case, we also fit the photometry with $\log g = 7, 9$ dex, and add the difference in measured radius and temperature in quadrature to the radius and temperature errors. For Gaia photometric fits, we always fit with fixed $\log g = 8$ dex because Gaia photometry offers only two independent constraints on the WD physical parameters, with $G$-band magnitudes being largely degenerate with $G_{BP}$ and $G_{RP}$ photometry. We find the photometric fits to be largely insensitive to the effects of fixing the surface gravity, and this is discussed in depth in our companion paper (N. R. Crumpler et al. 2024, in preparation). We perform these fits for the photometry corresponding to the 16th percentile, median, and 84th percentile C. A. L. Bailer-Jones et al. (2021) distances. We take the median distance measurements as the measured radius and temperature of the star. We add the difference in temperature or radius for the near and far WD distances in quadrature to the returned measurement errors to account for the uncertainty in the distance to the WD. We validate this fitting method by comparing our measured Gaia radii and temperatures to those from the N. P. Gentile Fusillo et al. (2021) catalog, which also uses Gaia photometry. We find agreement to within $0.0005 R_\odot$ and 3% for our radius and temperature measurements, respectively. We validate our SDSS measurements by comparing to our Gaia measurements, and find agreement to $0.001 R_\odot$ and 6% for our radius and temperature measurements. These SDSS photometric measurement systematic uncertainties are likely overestimates due to differences in using SDSS and Gaia photometry. Because the SDSS $u$-band constraint becomes important for WDs at high effective temperatures, we prioritize the use of SDSS over Gaia photometry despite the larger systematic errors. We fit both SDSS and Gaia photometry whenever available, fitting SDSS photometry to 21,766 WDs and Gaia photometry to 26,040 WDs. In our sample for Sections 3–4, we choose to use SDSS photometry if available, so we use Gaia photometry for only 4275 WDs.

## 3. Empirical White Dwarf Mass–Radius Relation

We use the catalog of Section 2 to characterize the WD mass–radius relation by isolating the gravitational redshift of the sample as a function of photometric radius or spectroscopic surface gravity. We find good agreement between our empirical mass–radius relation and a theoretical relation from the La Plata models.[16] These models contain tables of DA WD masses and radii as a function of effective temperature and surface gravity. For low-mass helium core WDs, intermediate-mass carbon-oxygen core WDs, and high-mass oxygen-neon core WDs, these models use the results of L. G. Althaus et al. (2013), M. E. Camisassa et al. (2016), and M. E. Camisassa et al. (2019), respectively. In the La Plata models, the thickness of the WD hydrogen envelope varies with the mass of the object, with thicker envelopes ($M_H/M_{WD} \sim 10^{-3.5}$) at lower masses and thinner envelopes at higher masses ($M_H/M_{WD} \sim 10^{-5.7}$). The ranges of hydrogen mass fractions included in this set of La Plata models are all considered to be thick hydrogen layers.

### 3.1. Quality Cuts and Final Sample Selection

To quantify the WD mass–radius relation, we use the measured radial velocities and surface gravities from the coadded spectrum associated with each SDSS-V or previous SDSS object. To maintain as many objects as possible while reducing noise in the data, we perform data quality cuts. As discussed in Section 2, spectroscopic parameters are best measured from spectra with an SNR of at least 10. So, we require the coadded spectrum SNR to be >10. Additionally, the P. E. Tremblay et al. (2013) models cover a grid in effective temperature ranging from $1500 < T_{eff} < 130{,}000$ K and in surface gravity ranging from $7 < \log g < 9$ dex. To remove objects that hit the edge of the grid during the fitting process, we remove any objects for which the coadded spectroscopic temperature or photometric temperature do not fall within the range of $1600 < T_{eff} < 129{,}000$ K. Similarly, we remove any WDs for which the coadded spectroscopic surface gravity falls on or outside the valid range of the P. E. Tremblay et al. (2013) models. Additionally, we remove any WDs for which the full radial velocity error is >50 km s$^{-1}$, the full photometric radius error is >0.006 $R_\odot$, or the full coadded spectroscopic surface gravity error is >0.3 dex. Finally, we select only those objects for which the median distance from C. A. L. Bailer-Jones et al. (2021) is <500 pc and remove WDs that are likely part of the Milky Way halo rather than the disk stellar population. We describe the purpose of and calculation of these cuts in Section 3.2. The most stringent of these requirements is the SNR cut, as most SDSS WD spectra have a SNR ≲ 5. After performing these cuts, our initial catalog is reduced from 8545 and 19,257 objects to 2887 and 7242 for SDSS-V and previous SDSS WDs, respectively. In Sections 3 and 4, all references to the SDSS-V and previous SDSS catalogs refer to these smaller catalogs of more precise measurements.

To investigate the general agreement of the data with theoretical models of WD structure, we combine the SDSS-V and previous SDSS WD catalogs into one catalog of all WD measurements passing the data quality cuts, hereafter referred to as the combined catalog. There are 723 objects in the combined catalog with measurements from both SDSS-V and previous SDSS data. For these objects, we take the weighted mean of the SDSS-V and previous SDSS measurements to obtain one value for each unique WD.

---

[13] https://dustmaps.readthedocs.io/en/latest/index.html
[14] https://extinction.readthedocs.io/en/latest/
[15] https://mfouesneau.github.io/pyphot/

[16] http://evolgroup.fcaglp.unlp.edu.ar/TRACKS/newtables.html

### 3.2. Averaging to Isolate Gravitational Redshift

The apparent radial velocity has two components, the true radial velocity ($v_r$) and the gravitational redshift ($v_g$), such that

$$v_{\rm app} = v_r + v_g. \tag{3}$$

The gravitational redshift is a general relativistic effect that appears because of the high surface gravity of WDs. As photons climb out of the steep potential well surrounding the WD, they lose energy and escape at a longer wavelength (A. Einstein 1916). This shift can be related to an apparent recession velocity via

$$v_g = \frac{GM}{Rc}, \tag{4}$$

where $M$ is the mass and $R$ is the radius of the WD.

Within the solar neighborhood ($\lesssim 500$ pc), objects can be considered to be part of the locally comoving frame termed the local standard of rest (LSR; J. Binney & S. Tremaine 1987). We correct the SDSS wavelengths, which are calibrated relative to the solar system barycenter, to this frame by accounting for the solar system's velocity relative to the LSR (R. Schonrich et al. 2010),

$$(U, V, W) = (11.1, 12.24, 7.25) \text{ km s}^{-1}. \tag{5}$$

The magnitude of this correction is 18 km s$^{-1}$ or less. The mean LSR corrections for the SDSS-V and previous SDSS catalogs are $-3.3$ and $-3.1$ km s$^{-1}$, respectively, so this is a small effect relative to the magnitude of the WD gravitational redshifts (tens of km s$^{-1}$). SDSS-V and previous SDSS WDs have different spatial distributions due to differences in survey strategies that are discussed in detail in our companion paper (N. R. Crumpler et al. 2024, in preparation). Previous SDSS WDs were targeted serendipitously in observing runs designed to search for extragalactic quasars, which are located outside the plane of the Milky Way, while SDSS-V WDs are intentionally targeted as WDs. Thus, SDSS-V has significantly more WDs located in the Galactic plane ($b \sim 0$ deg) than previous generations. This, and other effects from the differences in WD targeting, lead to systematic differences between SDSS-V WDs and WDs from previous generations of SDSS. In the case of the LSR correction, at the mean Galactic longitudes of $l = 140$ and 151 deg for the SDSS-V and previous SDSS catalogs, respectively, the mean LSR corrections to the radial velocity in the plane ($b = 0$ deg) for SDSS-V and above the plane ($b = 50$ deg) for previous SDSS generations are $-0.1$ and $-3.0$ km s$^{-1}$. Thus, the difference in spatial distribution can result in differences in LSR corrections on the order of 2–3 km s$^{-1}$ between the catalogs.

After correcting the radial velocities to the LSR, we can remove halo contaminants in our DA WD population. Halo stars have high radial velocities relative to the LSR. A selection cut requiring stars to have radial velocities within 120 km s$^{-1}$ of the LSR results in a halo contamination fraction $\leqslant 10\%$ (S. Ninkovic et al. 2012). We subtract the mean gravitational redshifts of our SDSS-V and previous SDSS samples, 31.9 and 27.0 km s$^{-1}$, respectively, from the radial velocities to obtain only the radial velocity relative to the LSR. We then require this radial velocity to be $<120$ km s$^{-1}$, removing 1.6% and 1.2% of our SDSS-V and previous SDSS samples, respectively. This large difference in mean gravitational redshift between the samples as well as other systematic differences between the SDSS-V and previous SDSS catalogs are investigated in N. R. Crumpler et al. (2024, in preparation). By comparing radial velocities for the same WDs observed in SDSS-V and previous generations, we find evidence that radial velocities are systematically larger in SDSS-V in our companion paper.

WDs are an older stellar population, and so they experience an asymmetric drift relative to the LSR (J. Binney & S. Tremaine 1987). For WDs in the SDSS-V catalog, the mean radial velocity of WDs located within 10° of $l = 270°$ is 25 km s$^{-1}$ larger than the mean radial velocity of WDs located within 10° of $l = 90°$. For the previous SDSS catalog, this signature of asymmetric drift is less pronounced, with WDs near $l = 270°$ having a mean radial velocity 11 km s$^{-1}$ greater than WDs near $l = 90°$. We calculate the velocity dispersion of the LSR-corrected radial velocities for the SDSS-V and previous SDSS WD catalogs. The velocity dispersions for the SDSS-V and previous SDSS WDs passing the data quality cuts of Section 3.1 are 37 and 34 km s$^{-1}$, respectively. From Figure 3 of R. Schonrich et al. (2010), radial velocity dispersions of 37 and 34 km s$^{-1}$ correspond to asymmetric drifts of magnitude 10 and 8 km s$^{-1}$ for the SDSS-V and previous SDSS WDs, respectively. Due to the asymmetric drift, WDs rotate more slowly than the LSR about the Galactic Center and so this asymmetric drift is a negative contribution to the $V$ component of a WD's $(U, V, W)$ Galactic Cartesian coordinate velocity, which we then project onto the LSR and add to the WD LSR-corrected radial velocities of each WD. These asymmetric-drift corrections are 14 km s$^{-1}$ or less in magnitude. This correction largely cancels during the averaging process to isolate gravitational redshifts, with the mean asymmetric drift corrections for the SDSS-V and previous SDSS catalogs being 1.0 and 2.1 km s$^{-1}$.

After these corrections, the apparent radial velocities are the sum of the WDs' gravitational redshifts and the randomly distributed radial velocities of the WDs from noncircular orbits around the Galactic center. These random motions can be averaged out by binning the sample in photometric radius or in spectroscopic surface gravity, leaving only the gravitational redshift contribution (R. E. Falcon et al. 2010; V. Chandra et al. 2020b).

For each of the SDSS-V, previous SDSS, and combined catalogs, we bin the LSR and asymmetric drift-corrected coadded radial velocity measurements in photometric radius and spectroscopic surface gravity. The size of these bins is determined by the typical uncertainty of our radius and surface gravity measurements so that we can reasonably assume most WDs fall into the correct bin. The radius bins range from 0–0.054 $R_\odot$, and the width of each bin is 0.002 $R_\odot$. The surface gravity bins range from 7–9 dex, and the width of each bin is 0.1 dex. Within each radius or surface gravity bin, for each catalog, we take the weighted mean of the corresponding radii or surface gravities and radial velocities. The mean radial velocity of each bin is the gravitational redshift for the WDs in the bin. To estimate the error on each of the weighted means, we draw a sample from each of our catalogs with replacement of size equal to the size of our catalog and recompute the sample weighted means. We repeat this bootstrapping process for 1000 samples and calculate the standard deviation of the sample means. We add this standard deviation in quadrature to the error on our measured weighted means to obtain the full error on the weighted means. The gravitational redshift and either radius or surface gravity can be combined to obtain the mass of the WDs in that bin via

$$M(R, v_g) = \frac{v_g R c}{G}, \quad (6)$$

$$M(\log g, v_g) = \frac{v_g^2 c^2}{G 10^{\log g}}. \quad (7)$$

Thus, this binning, averaging, and bootstrapping routine allows us to obtain the gravitational redshift and mass as a function of photometric radius and spectroscopic surface gravity.

### 3.3. Bias Corrections

For each catalog of WD observations, the binning, averaging, and bootstrapping routine of Section 3.2 gives the gravitational redshift and mass as a function of photometric radius and spectroscopic surface gravity. But these averaged parameters are biased due to the sharply peaked distributions of WD surface gravities and radii. Because of measurement noise, a WD near the peak of the distribution is likely to fall into a nearby bin, flattening the distribution. For catalogs binned in surface gravity, this effect artificially increases the measured gravitational redshifts of $\log g$ bins $<7.95$ dex and decreases the gravitational redshifts of bins $>7.95$ dex as objects with $\log g \sim 8$ fall on either side of the distribution. Similarly, for catalogs binned in radius, this effect artificially increases the measured gravitational redshifts of radius bins $>0.013\,R_\odot$ and decreases the gravitational redshifts of bins $<0.013\,R_\odot$.

We create a Monte Carlo simulation to characterize this bias due to binning noisy measurements from sharply peaked distributions. For a given catalog, we draw a truth sample of the same size with masses from the mass distribution given in Table 1 of S. O. Kepler et al. (2007) and temperatures drawn from the photometric temperatures measured in our catalog. We use the theoretical relations from the La Plata models to obtain the radii, surface gravities, and gravitational redshifts for this truth sample. Then, we add Gaussian noise to the radius and surface gravity samples, resulting in a noisy radius and noisy surface gravity sample. We run one simulation in which the width of the Gaussian noise is given by the median full error of our measured radii or surface gravities. This median error is $0.0014\,R_\odot$ and $0.12$ dex for the radii and surface gravities included in the combined catalog. We also run a low and high noise simulation to characterize the systematic uncertainty on our weighted mean parameters due to this bias correction. In the low noise simulation, the width of the noise is given by the median full error minus one standard deviation of the full errors. In the high noise simulation, the noise is given by the median error plus one standard deviation. The standard deviations of the full errors on the radii and surface gravities measured in the combined catalog are $0.0011\,R_\odot$ and $0.04$ dex, respectively. We bin the truth and noisy samples, and calculate the effect on the measured gravitational redshift. The bias due to this effect is the difference between the truth and the noisy samples. We repeat this process for 10,000 noise realizations, and average the resulting bias corrections. Figure 1 shows the output of one of these realizations.

We run this simulation 10 times for each of the SDSS-V, previous SDSS, and combined catalogs. In each run, we add the resulting bias corrections from the median noise simulation to the measured gravitational redshifts, masses, mean radii, and mean surface gravities from Section 3.2. We take the difference in the bias corrections for each parameter from the low and high noise simulations, and add this systematic uncertainty in quadrature to the uncertainty on the weighted mean parameters.

The uncertainties on the mean surface gravities are dominated by measurement uncertainties; uncertainties on the mean radii are dominated by the error on the bias correction at small radii ($<0.02\,R_\odot$) and by measurement uncertainties at large radii ($>0.02\,R_\odot$); and the uncertainties on the gravitational redshifts are dominated by the error on the bias correction at small radii ($<0.02\,R_\odot$) and intermediate surface gravities ($8.0 < \log g < 8.5$ dex) and by measurement uncertainties in the rest of the radius or surface gravity parameter space. For all bias-corrected parameters, we verify that all 10 bias simulation runs produce the same results.

### 3.4. Effects of Binary Contamination

A potential source of scatter in our measurements comes from the effects of binaries in our sample. There are two main classes of binaries that could contaminate the sample, double WD (DWD) binaries and WD-main-sequence star (WDMS) binaries. These binaries might be identified by characterizing the shift in measured radial velocity among multiple observations or within the sub-exposures of a single observation of a given WD (G. Adamane Pallathadka et al. 2024, in preparation). We do not try to remove these binaries and restrict our analyses to the regime in which the effects of binary contamination are negligible.

For DWD binaries, we determine the regime in which these effects are negligible by simulating the effect of their contamination on our measured gravitational redshifts, masses, mean radii, and mean surface gravities. As in the sharp-peak bias correction, we draw a no-binary sample of the same size as our combined catalog with masses from the mass distribution given in Table 1 of S. O. Kepler et al. (2007) and temperatures drawn from the photometric temperatures measured in our catalog. As before, we obtain the radii, surface gravities, and gravitational redshifts for this no-binary sample. Then we create a corresponding DWD binary sample. We set the parameters of the no-binary sample to be the properties of the primary WDs in the binary sample. For each primary WD, if the WD mass is $<0.45\,M_\odot$, we set the probability of having a binary companion equal to 1 (T. R. Marsh et al. 1995). At such low masses, WDs must have evolved via binary evolution in order to become a WD over the age of the Universe. If the primary mass is greater than this threshold, we give the star a 5% chance of having a binary companion. Five percent is chosen because the overall unresolved WD binary percentage found in the literature varies from 1%–5% (J. B. Holberg 2009; S. Toonen et al. 2017; S. Torres et al. 2022), although for systems with small separations ($<4$ au), this percentage can be as high as 10% (D. Maoz et al. 2018). Overall, this results in a binary sample with a total binary proportion of $\sim13\%$. This represents a worst-case scenario since the real binary proportion is likely far less than this.

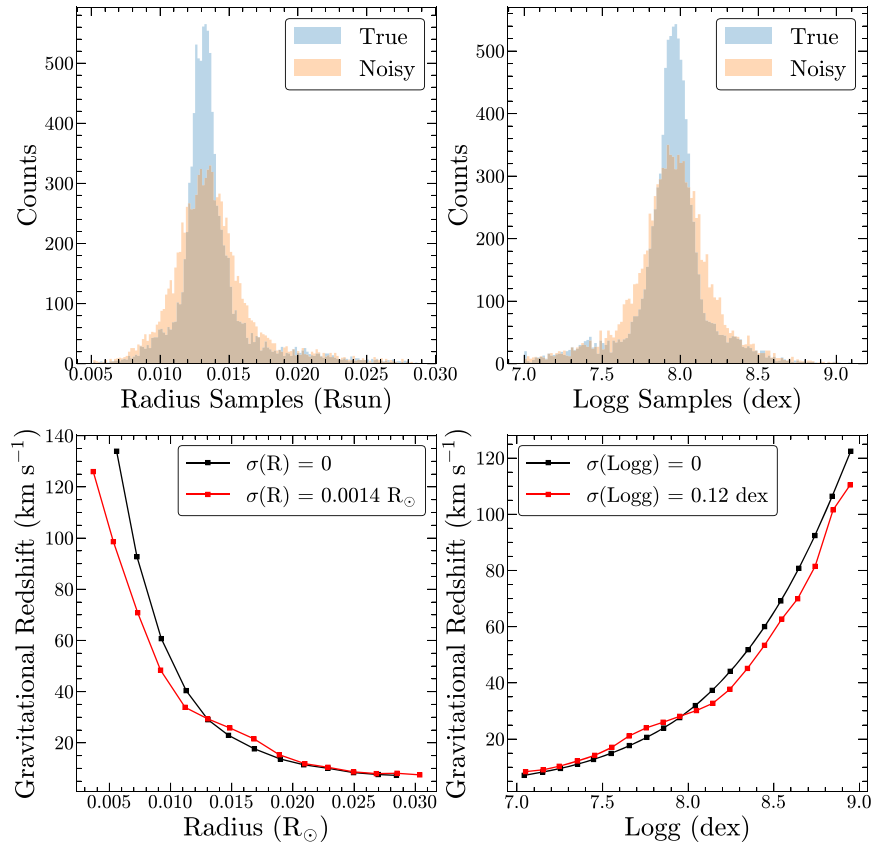

**Figure 1.** Example run of the sharply peaked distribution bias simulation. The top plots show the true (blue) and noisy (orange) radius (left) and surface gravity (right) distributions. Adding noise to the true distributions flattens them. The bottom plots show the effect of this flattening on measuring gravitational redshifts. The relation between the true radius or surface gravity and the gravitational redshift is shown in black, and the noisy relation is shown in red.

For each DWD binary system, we draw a companion mass from our primary mass sample, requiring the companion to be more massive and thus smaller than the primary WD. We also draw a temperature from our primary temperature sample and compute the companion radius, surface gravity, and gravitational redshift. We draw a system orbital separation from the D. Maoz et al. (2018) distribution and use it to compute the period of the system as well as the orbital velocities of both the primary and companion assuming circular orbits. We draw a line-of-sight inclination from the D. Maoz et al. (2012) distribution, choose a random orbital phase, and calculate the primary and companion radial velocities. We then trim the sample so that all WDs have surface gravities and effective temperatures in the ranges covered by the P. E. Tremblay et al. (2013) models. We then use the P. E. Tremblay et al. (2013) models to build model spectra for both the primary and companion WD in each binary system. We add these spectra to obtain the model spectrum for the full binary system. Also, we obtain model SDSS $urz$ and Gaia $G_{BP}$ and $G_{RP}$ photometry for the primary and companion using the G. Fontaine et al. (2001) interpolation from mass and effective temperature to absolute magnitudes implemented in the publicly available package `WDmodels`.[17] We add these model magnitudes in flux space and convert them back to magnitudes to obtain the SDSS or Gaia photometry for the whole binary system.

We then repeat the measurement procedures of Section 2, given the model spectra and photometry of the DWD binary systems. We repeat the averaging procedure of Section 3.2 for the samples with and without binaries, and compare the results. Overall, even with a higher binary proportion than estimated in the literature, the effect of binaries on our measured gravitational redshifts is negligible ($\ll 1\,\mathrm{km\,s^{-1}}$) so long as the mass of the WD is $>0.45\,M_\odot$. This is because, at sufficiently high masses, the binary contamination is small enough that averaging over many WDs is robust against this source of contamination, even if the presence of binaries does add to the RV dispersion of the sample.

For masses $<0.45\,M_\odot$, nearly all WDs can be assumed to be in a binary system, and DWD binaries can affect the measured mass–radius relation in this regime. In this low-mass regime, the gravitational redshifts computed using radial velocities binned in photometric radius are $\sim$10–20 km s$^{-1}$ higher than the true values. This moderate increase is due to the fact that the photometric radius measurements are not dramatically impacted by the presence of an additional WD companion, with the mean measured binary system radius being 0.002 $R_\odot$ larger than the primary radius. This increase is of the same order of magnitude as the uncertainties on the radius measurements and the size of the radius bins. In contrast, the gravitational redshift measurements for binary systems are biased 10–20 km s$^{-1}$ higher than the primary gravitational redshift. This is because the gravitational redshifts of binary systems are computed as the mean gravitational redshift of the primary and companion plus the measured system radial velocity. Because we require the companion to be more massive than the primary, this increases the observed gravitational redshift even if the positive and negative contributions of the measured radial velocities roughly cancel. This is a worst-case scenario for the bias effects due to DWD binaries, as we typically do not observe the secondary in practice.

---
[17] https://github.com/SihaoCheng/WD_models

For gravitational redshifts computed using radial velocities binned in spectroscopic surface gravity, the effect of DWD binary systems is more pronounced in the low-mass ($<0.45\,M_\odot$) regime. The presence of a WD binary companion broadens the spectral lines, causing the fit to record a higher surface gravity than the true value. The mean measured binary system surface gravity is 0.3–0.4 dex larger than the primary surface gravity. This increase is much larger than the size of the surface gravity bins (0.1 dex). Thus, these binary systems become contaminants in higher surface gravity bins. In the high surface gravity bins, the averaging procedure is robust against the effects of these contaminants. However, there are very few WDs remaining in the low-mass regime, increasing the noise and decreasing the efficacy of the averaging procedure to measure gravitational redshifts.

Overall, the results of the simulation suggest that our averaging routine is only valid for masses $>0.45\,M_\odot$, which corresponds to radii $<0.016\,R_\odot$ or log surface gravities $>7.7$ dex for a 10,000 K WD. Thus, we restrict ourselves to this sufficiently high-mass regime when investigating WD structure in Sections 3.6 and 4. We verify these conclusions by repeating the results of Sections 3.6 and 4 removing SDSS-V objects flagged as potential DWD binaries by G. Adamane Pallathadka et al. (2024, in preparation). Of the 152 objects listed as likely DWD binaries, 120 are in our full SDSS-V DA WD catalog, and 54 are in the subset of the SDSS-V catalog passing the data quality cuts of Section 3.1. This corresponds to contamination fractions of 1.4% and 1.8% for the full SDSS-V catalog and the subset passing quality cuts, respectively. This contamination is far less than the fraction used in our DWD binary simulation, which showed no effect on our results so long as $M_\mathrm{WD} > 0.45\,M_\odot$, even for a contamination fraction of ~13%. Removing these 120 objects does not affect the results of Sections 3.6 and 4.

On average, WDs in WDMS binaries are much fainter than their main-sequence companions. Because of this, many WDMS binaries are unresolved and classified instead as isolated main-sequence stars. Such systems do not contaminate our catalogs of DA WDs. For those WDMS binaries in which the WD can be observed, most fall above the WD sequence on the color–magnitude diagram due to the presence of the brighter companion and thus can be readily identified (A. Rebassa-Mansergas et al. 2021). The SDSS WD targeting strategy is designed to avoid these obvious binaries, although there is a possibility of unresolved or partially resolved WDMS binaries contaminating the WD sample. These unresolved WDMS systems consist of WDs with dim main-sequence-star companions, typically M-dwarfs. There are a few known examples of these systems in which the M-dwarf is small enough and the WD is hot enough that the WD completely dominates the optical and UV spectrum (T. R. Marsh & S. R. Duck 1996; P. F. L. Maxted et al. 1998; C. Badenes et al. 2013). But most of these WD and M-dwarf systems can be identified by an excess in the red part of the object's spectrum compared to an isolated WD (A. Rebassa-Mansergas et al. 2009). The `SnowWhite` classifier has been trained on a couple thousand such binaries, and is able to identify these systems with red excesses as WDMS binaries. We do not include objects classified as binaries by `SnowWhite` in our sample, and thus the level of WDMS contamination in our sample is low. In an ~80% complete 100 pc volume-limited sample of unresolved WDMS binaries in which a WD component could be identified, only 112 systems were discovered. This is very few compared to nearly 13,000 WDs contained in the comparable ~95% complete 100 pc volume-limited sample of Gaia WDs (A. Rebassa-Mansergas et al. 2021). This suggests a contamination rate of <1% among the general WD population. Given that we remove systems classified as unresolved WDMS binaries by `SnowWhite`, our contamination rate should be ≪1%. The results of the DWD binary simulation show that the averaging procedure is generically robust against this very low level of contamination. Thus, these WDMS binary contaminants can be neglected in the results of Sections 3.6 and 4.

### 3.5. Effects of Thin Hydrogen Layer White Dwarf Contamination

The thickness of the hydrogen layer enveloping DA WDs is currently poorly constrained (P. E. Tremblay et al. 2013). Many models assume constant thick hydrogen layers with $M_\mathrm{H}/M_\mathrm{WD} \sim 10^{-4}$ (I. J. Iben & A. V. Tutukov 1984). There is evidence that this thickness varies among WDs, with some WDs having thin hydrogen layers for which the hydrogen mass fraction is $M_\mathrm{H}/M_\mathrm{WD} \sim 10^{-10}$ (G. Fontaine et al. 2001). Although the number of thin layer DA WDs is thought to be far less than those with thick hydrogen layers (≲20%; P. E. Tremblay & P. Bergeron 2008), the impact of the hydrogen layer thickness has been found to have important consequences for some methods of empirically testing the WD mass–radius relation (A. D. Romero et al. 2019). In particular, the measured radius and surface gravity of a WD are dependent on the assumption of thick or thin hydrogen layers. Radial velocity measurements and, by extension, gravitational redshifts are robust against the assumed hydrogen layer thickness, since these measurements come from the centroid and not the shape of the Balmer series lines. Both the P. E. Tremblay et al. (2013) and La Plata models employed in this paper utilize thick hydrogen layers.

We perform a simulation to characterize the importance of contamination by thin hydrogen layer WDs in our sample. As a conservative estimate, we assume the highest contamination fraction indicated by the literature, randomly selecting 20% of our sample to be thin hydrogen layer WDs. For each selected WD, we correct the measured surface gravity and radius to values assuming the object instead has a thin hydrogen layer. We obtain these corrections by taking the difference in the radius or surface gravity predicted by the thin and thick hydrogen layer models of G. Fontaine et al. (2001) at the measured photometric temperature of the WD, fixing the WD mass to be near the peak of the WD mass distribution of $0.6\,M_\odot$. The median differences between thin and thick layer surface gravities and radii are 0.037 dex and $-0.0006\,R_\odot$, respectively, such that thin hydrogen layer WDs tend to have larger surface gravities and smaller radii. But these differences can be as large as 0.1 dex and $0.0019\,R_\odot$ in magnitude. Thus, the size of these corrections can be comparable to the bin size used in our averaging procedure, meaning that WDs can fall into nearby radius or surface gravity bins due to this effect. Our measured radial velocities are not significantly affected by the assumed thickness of the hydrogen layer surrounding the WD. We then repeat the binning and averaging routine of

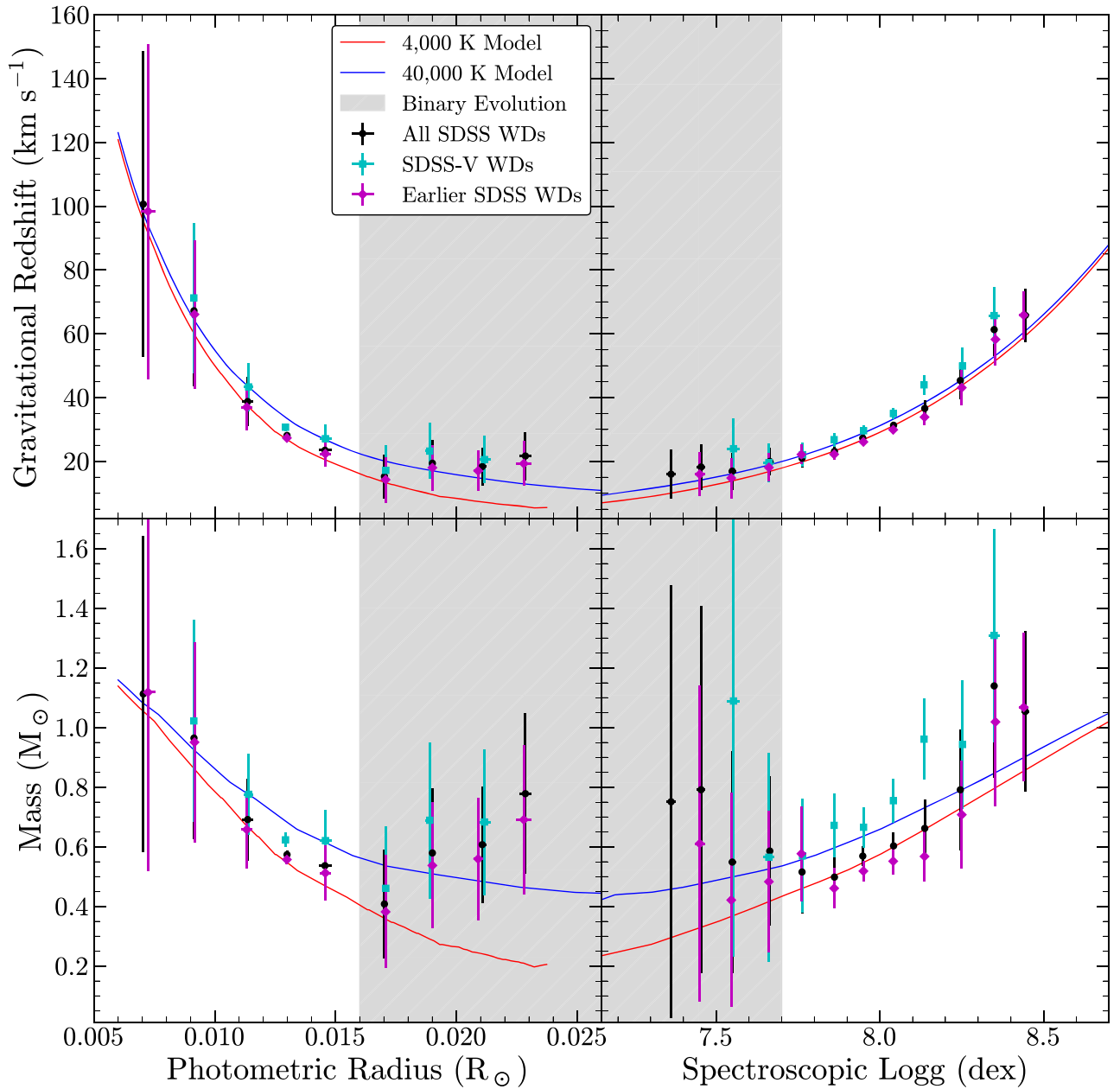

**Figure 2.** The agreement between the La Plata theoretical models of WD structure and our measurements. The core composition and hydrogen layer thickness of the La Plata models varies as a function of WD mass. The model for a very cold 4000 K WD is shown in red and for a hot 40,000 K WD is shown in blue. The gravitational redshifts (top) are binned in photometric radius (left) and in spectroscopic surface gravity (right). In the bottom row, we plot the corresponding mass–radius or mass–surface gravity relations for the theoretical models and our data. We show our measurements using only the SDSS-V catalog in cyan, only the previous SDSS catalog in magenta, and the combined catalog in black. The error bars on the data come from the error on the weighted mean radial velocity plus the systematic error due to the sharp-peak bias correction, added in quadrature. The region in which most WDs must be in binary systems is shaded in gray. We find that our empirical mass–radius relation is consistent with the La Plata models.

Section 3.2 for this contaminated sample. We recreate all results of the following Sections with this contaminated sample, and find no significant effect on our measurements. The effects of any variation due to contamination by thin hydrogen layer WDs are subdominant to the effects from the uncertainty in the sharp-peak bias correction of Section 3.3.

### 3.6. White Dwarf Structure Measurement Results

In Figure 2, we show the agreement between the La Plata theoretical models, which range in core composition and hydrogen layer thickness, and the bias-corrected gravitational redshifts computed from the LSR and asymmetric drift-corrected radial velocities from SDSS-V and previous SDSS coadded spectra. This plot corresponds to one run of the bias-correction simulation, and all other implementations of the bias correction show similar results. We show the agreement for gravitational redshifts binned in photometric radius and binned in spectroscopic surface gravity. We emphasize that, in this method, the measurements of model spectra need only be accurate to within the bin width of the averaging procedure. Thus, this averaging procedure is robust against the underlying assumptions of model spectra. Neglecting the regime in which most WDs are in binaries, our results are consistent with the La Plata theoretical models. For the combined and previous SDSS catalogs, nearly all gravitational redshift measurements are within $1\sigma$ of the 4000 and 40,000 K theoretical relations. For the SDSS-V catalog, all gravitational redshifts measured by binning in photometric radius are also within $1\sigma$ of the theory. For the SDSS-V redshifts measured from spectroscopic surface gravity, some of the SDSS-V values are systematically higher than the La Plata curves. Overall, we find that SDSS-V WDs have systematically larger radial velocities than WDs from previous generations of SDSS, potentially accounting for this offset. We discuss the difference in radial velocity distributions between SDSS-V and previous SDSS observations in depth in our VAC companion paper, N. R. Crumpler et al. (2024, in preparation).

The sharp-peak bias correction of Section 3.3 is minimized near the peaks of the WD radius or surface gravity distributions at $\sim 0.013\,R_\odot$ and $\sim 7.95$ dex, respectively. Far from these peaks, the bias correction becomes strong, and our results are very sensitive to the median full error on our radius and surface gravity measurements. Thus, the overall agreement of our measurements with the La Plata theoretical values is best characterized near these peaks, where our results are most robust against the effects of bias corrections and where we have the most data available. In Figure 3, we show the agreement between our measured gravitational redshift at the peak of the radius and surface gravity distributions and the theoretical gravitational redshift from the La Plata models for one run of the bias correction simulation. We select 4245 WDs with radii within the range of $[0.012, 0.014]\,R_\odot$ and 2445 WDs with surface gravities within the range of $[7.9, 8.0]$ dex. The other nine bias correction runs show the same result. The error on our gravitational redshift comes from the measured error on the weighted mean combined with the bootstrapped uncertainty on the weighted mean and the error due to the bias correction. We calculate the La Plata theoretical gravitational redshift at the peak radius or peak surface gravity and the mean photometric effective temperature of the WDs in the peak. The mean effective temperature of the WDs in the radius peak is 14,300 K and of the WDs in the surface gravity peak is 14,000 K. The error on the theoretical gravitational redshift is given by computing this theoretical redshift at $\pm 1\sigma$ in photometric temperature. The mean measured and theoretical gravitational redshifts for the peak radius bin are $28.1 \pm 0.6$ and $29.5^{+1.0}_{-1.2}$ km s$^{-1}$, respectively. The mean measured and theoretical gravitational redshifts for the peak surface gravity bin are $27.3 \pm 0.8$ and $27.5^{+0.3}_{-0.3}$ km s$^{-1}$, respectively. Both our measurements binned in radius and surface gravity agree with the theoretical models to within the stated uncertainties.

## 4. Temperature Dependence of White Dwarf Structure

### 4.1. Current Measurements from the Available SDSS Data

We use the catalogs of Section 3 to directly detect the temperature dependence of the WD mass–radius relation. To investigate the dependence of WD structure on temperature, we divide the combined SDSS-V and previous SDSS catalog into cool and warm WD subsamples. The cool subsample consists of WDs with photometric effective temperatures <11,000 K and the warm sample consists of WDs with temperatures >16,000 K. These thresholds were chosen so that the combined catalog would contain roughly equal numbers of WDs in each temperature range. The combined catalog subsamples contain

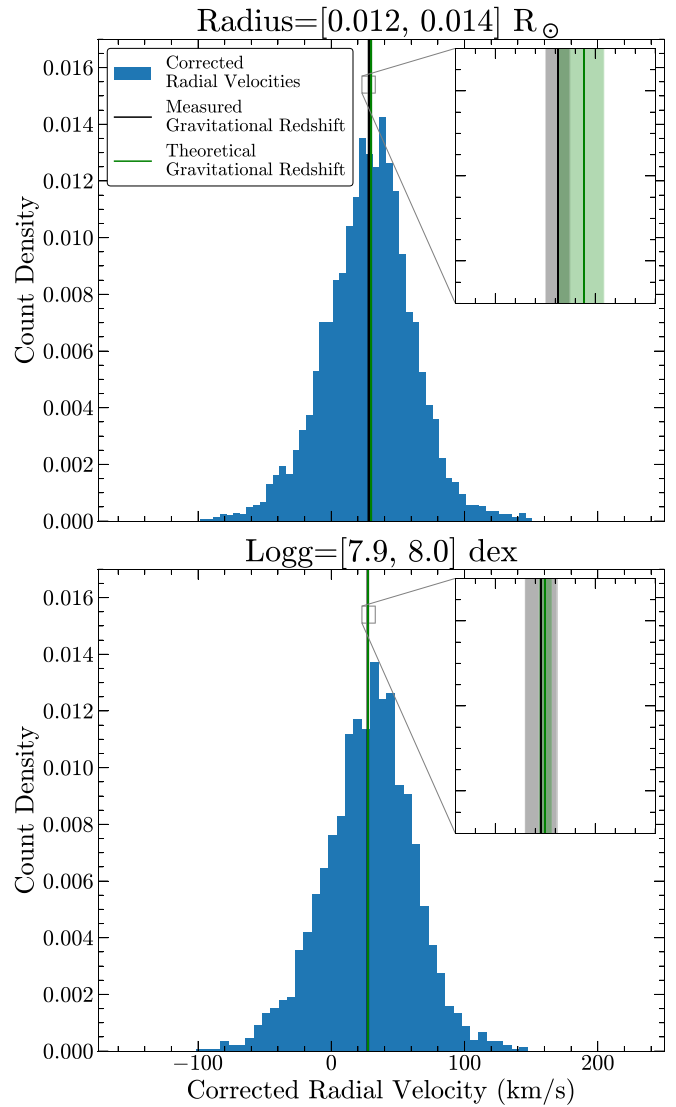

**Figure 3.** The agreement between our measured gravitational redshifts and the theoretical redshifts from La Plata models at the peak of the radius (top) and surface gravity (bottom) distributions with no temperature cuts. Our measured gravitational redshift, including bias corrections, is shown in black, with the gray shaded region showing the error on this measurement. The theoretical gravitational redshift at the peak radius or peak surface gravity and the mean photometric effective temperature of the WDs in the peak is shown in green. The uncertainty on this theoretical gravitational redshift is the green shaded region. The mean measured and theoretical gravitational redshifts for the peak radius bin are $28.1 \pm 0.6$ and $29.5^{+1.0}_{-1.2}$ km s$^{-1}$, respectively. The mean measured and theoretical gravitational redshifts for the peak surface gravity bin are $27.3 \pm 0.8$ and $27.5^{+0.3}_{-0.3}$ km s$^{-1}$, respectively. The LSR and asymmetric drift-corrected radial velocities in the peak bin are shown in the blue histogram. We cannot bias-correct individual radial velocities, so the maximum of this histogram does not correspond precisely to the measured gravitational redshift.

3493 cool and 3394 warm WDs. Note that 1059 of the cool WDs are contained in the SDSS-V catalog, and 2676 are contained in the previous SDSS catalog. Additionally, 898 of the warm WDs are contained in the SDSS-V catalog, and 2783 are contained in the previous SDSS catalog. Figure 4 shows the distribution of the photometric effective temperatures within the cool and warm subsamples. The mean photometric temperatures of the cool and warm subsamples are 8500 and 21,500 K, respectively. Ninety-five percent of WDs in the cool sample are hotter than 6400 K, and 95% of WDs in the warm sample are cooler than 32,800 K. These temperature cuts exclude 2519 WDs with temperatures between 11,000 and 16,000 K. We repeat the binning, averaging, and bootstrapping routine of Section 3.2 and the bias-correction routine of Section 3.3 for each of the combined catalog cool and warm subsamples.

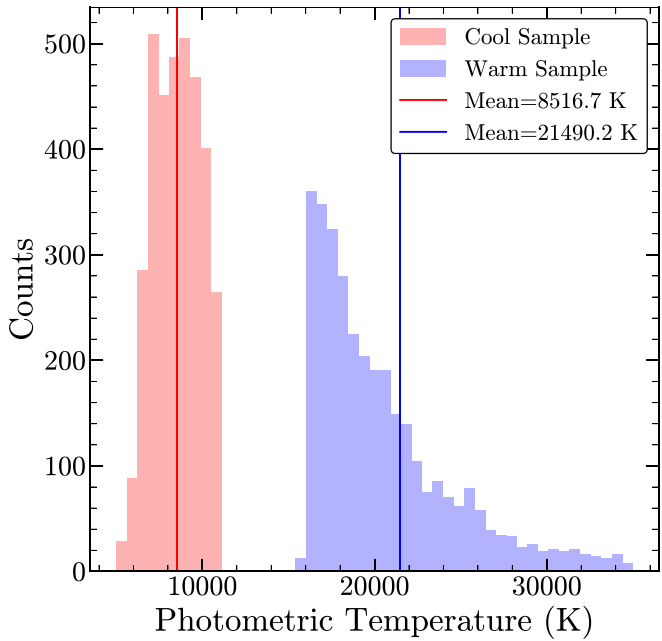

**Figure 4.** The distribution of WD photometric effective temperatures for the cool (red) and warm (blue) subsamples. The mean photometric temperatures of the cool (red line) and warm (blue line) subsamples are 8500 and 21,500 K, respectively.

Figure 5 is the same as Figure 2, but divided by photometric temperature into cool (<11,000 K) and warm (>16,000 K) subsamples for the combined catalog. The other nine implementations of the bias correction simulation show similar results. We find that, for gravitational redshifts measured both by binning in photometric radius or in spectroscopic surface gravity, cooler WDs systematically have smaller gravitational redshifts, as expected from the theoretical models. The photometric radius results are especially compelling, since these results are less reliant on the assumptions underlying model spectra than the results from surface gravity measurements. We emphasize that the La Plata models are plotted as a function of fixed radius or surface gravity, not fixed mass. At fixed mass, a hotter WD has a larger radius or smaller surface gravity, and thus a smaller gravitational redshift. At fixed radius or surface gravity, a hotter WD is more massive and thus has a larger gravitational redshift, as is shown in Figure 5.

Although the agreement between our measurements and theory in Figure 2 for our full sample is within the stated errors, in Figure 5 we see some disagreement between the theoretical La Plata models and our results. For gravitational redshifts computed by binning in photometric radius, there is one point in the cool sample for which our measured gravitational redshift falls below the 4000 K theoretical model. For gravitational redshifts computed by binning in surface gravity, there are four points in the cool sample for which our measured gravitational redshifts fall below the 4000 K theoretical model. There are also two points for which our measured gravitational redshifts fall above the 40,000 K theoretical model. We attribute these differences to a combination of different factors.

Both the results using photometric radii and spectroscopic surface gravities show very preliminary evidence that our mass measurements of cool WDs could be smaller than predicted by theory. There are some effects unique to cool WDs that could contribute to this disagreement. As the effective temperature of hydrogen-dominated WDs decreases, the strength of the Balmer lines also decreases until, at ∼5000 K, the lines disappear altogether (D. Saumon et al. 2022). Thus, spectroscopic fits become more difficult at lower effective temperatures, especially for spectra with poor SNRs. At low effective temperatures ($\lesssim$14,000 K), convection becomes increasingly important in characterizing the spectrum of WDs (P. E. Tremblay et al. 2013). This effect can reduce the reliability of DA WD spectroscopic models at low temperatures. Additionally, at low effective temperatures, magnetism becomes more common among WDs. For isolated WDs, this is potentially due to the generation of a crystallization-induced dynamo as the WD cools (J. Isern et al. 2017). Zeeman splitting from the presence of moderate magnetic fields can broaden Balmer series lines. Because at low temperatures these lines tend to be weaker, magnetic fields may have a more noticeable impact on Balmer line shapes in cool WDs than in higher-temperature WDs, thus disrupting spectroscopic fits at low effective temperatures. Low-temperature WDs are dimmer and thus harder to observe at sufficient SNR, resulting in selection effects. The culmination of these effects could explain part of the discrepancy between our measurements and the low-temperature La Plata models.

Other than a single point, the photometric radius results generally agree well with the La Plata models. Thus, the larger disagreements between our gravitational redshifts measured using spectroscopic surface gravities and the La Plata models are indicative that our surface gravity results are less reliable than our results from photometry. We see this also in Figure 2, although the surface gravity results are consistent with the La Plata models, the radius results show better agreement. Part of this discrepancy could be due to deeper uncertainties in the sharp-peak bias correction of Section 3.3. The effect of the bias correction is to decrease or increase the measured gravitational redshift for bins with surface gravities less than or greater than 7.95 dex, respectively. Comparing the bias-corrected redshifts to those without these corrections, we find that the bias corrections on the surface gravity measurements are much larger than those on the radius measurements. The median difference between the bias-corrected and uncorrected masses measured using photometric radii is −0.02 and 0.19 $M_\odot$ for the cool and warm samples, respectively. For masses measured using spectroscopic surface gravities, these differences are −0.04 and 0.21 $M_\odot$ for the cool and warm samples, respectively. Thus, the bias correction is more important for the cool WDs when using spectroscopic results. In the case of the lowest-mass data point, this bias correction pushes the measured mass down from 0.39 $M_\odot$ to 0.27 $M_\odot$, much farther from the mass predicted by theory. Our understanding of the effects of this sharp-peak bias correction is sufficient for the purpose of this paper, to detect the temperature dependence of the WD mass–radius relation. However, Figure 5 clearly shows that in order to push this technique further and use these methods to verify the agreement of empirical mass–radius relations as a function of temperature with theoretical models, it will be necessary to better characterize the systematic effects relating to the sharp-peak bias correction.

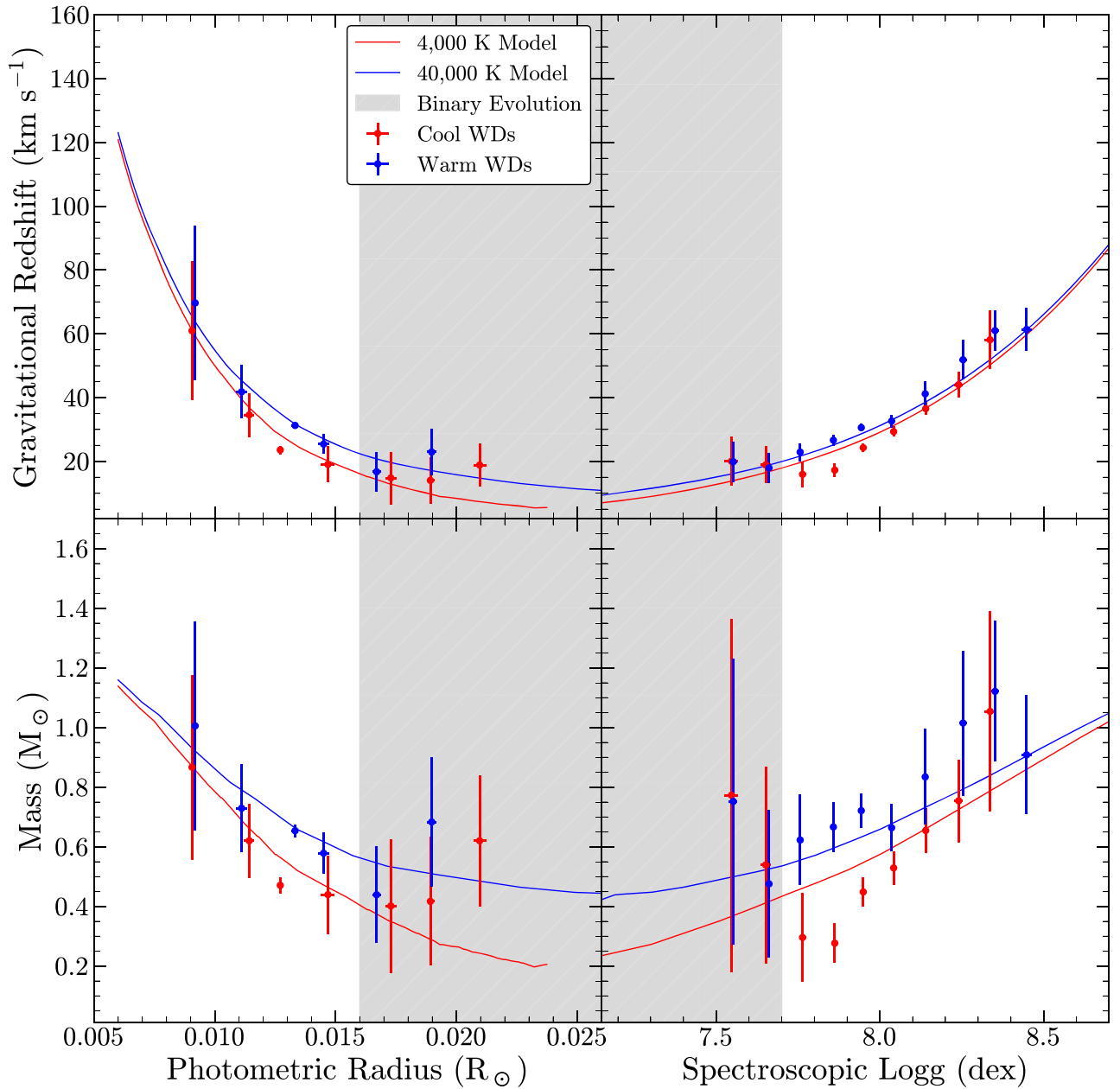

**Figure 5.** Same as Figure 2 but divided by temperature. We show our measurements using the combined catalog divided into a cool WD subsample (<11,000 K) in red and a warm WD subsample (>16,000 K) in blue. We find that cool WDs have smaller gravitational redshifts than warm WDs at a given radius or surface gravity, in line with our expectations from theory.

Figure 6 shows the distributions of measured photometric radii and spectroscopic surface gravities for the cool and warm subsamples of the combined catalog. This Figure does not rely on the bias corrections of Section 3.3, and the bins used in this Figure are the same as the bins used in Figures 2 and 5. Although the peaks of the cool and warm samples occur in the same radius or surface gravity bin, there are clear offsets in the distributions. These offsets can be characterized by calculating the weighted mean radius and surface gravity for the cool and warm subsamples. As in Section 3.2, we bootstrap these weighted means with 1000 samples to fully characterize the errors on these measurements. The mean radius for the cool and warm samples is $0.01250 \pm 0.00007$ and $0.01304 \pm 0.00008\,R_\odot$, respectively. The mean surface gravity for the cool and warm samples is $7.958 \pm 0.005$ and $7.918 \pm 0.004$ dex, respectively. We find that the mean radius of the warm sample is $5.2\sigma$ larger than that of the cool sample, and the mean surface gravity of the warm sample is $6.0\sigma$ smaller than that of the cool sample. The $\sigma$ values are computed as the difference in the two weighted means divided by each of their uncertainties added in quadrature. This is in line with the expectation from theory. Once an isolated WD of a given mass is formed, the star cools over time at constant mass. Thus, the peaks of the cool and warm subsamples should occur at roughly the same mass, and, at a fixed mass, a warmer WD has a larger radius and thus a smaller surface gravity.

To further characterize the significance of our measured temperature dependence of WD structure, we investigate the behavior of the mean gravitational redshifts at the peak of the measured radius ($\sim 0.013\,R_\odot$) and surface gravity ($\sim 7.95$ dex) distributions. At these peaks, we have the most data, and our results are less sensitive to the effects of the sharp-peak bias correction. Figure 7 shows the difference in measured gravitational redshifts between the cool and warm subsamples of the combined catalog for redshifts measured by binning at the peak radius and peak surface gravity for one run of the bias correction simulation. All other runs show similar results, although the significance of the temperature detection varies slightly. The peak of the radius distribution occurs in the $[0.012, 0.014]\ R_\odot$ bin. There are 1480 and 1572 WDs in this bin in the cool and warm subsamples, respectively. In Figure 7, we show the LSR and asymmetric drift-corrected radial velocities in this peak radius bin as well as the mean-radius-corrected gravitational redshifts of each subsample. To perform this mean radius correction, we calculate the mean of the cool and warm sample radius peaks and the mean temperature of each sample. We then use the La Plata models to correct each gravitational redshift to the same mean radius, at the measured temperature of the sample. This allows the gravitational redshifts from the cool and warm samples to be directly compared at the same radius value. We neglect any uncertainties on this correction. With this correction and for this bias-correction run, the mean redshift for the cool and warm samples computed by binning in radius is $22.0 \pm 1.3$ and $32.7 \pm 1.0$ km s$^{-1}$, respectively. This corresponds to a significance of the difference between the measured gravitational redshifts, computed as the difference in the two redshifts divided by each of their uncertainties added in quadrature, of $6.4\sigma$ with the sign predicted by theory. Before applying this correction, the significance of the difference for this bias-correction run is $4.6\sigma$ in the same direction, with cooler WDs having smaller gravitational redshifts at a fixed radius. Across all bias simulation runs, the significance of the mean-radius-corrected gravitational redshifts varies between $6.1\sigma$ and $6.5\sigma$ and of the uncorrected redshifts varies between $4.4\sigma$ and $4.6\sigma$. Thus, we find strong evidence for the temperature dependence of WD structure at the peak of the radius distribution, with a minimum significance of this detection of $6.1\sigma$ and a median significance of $6.4\sigma$ for the mean-radius-corrected redshifts.

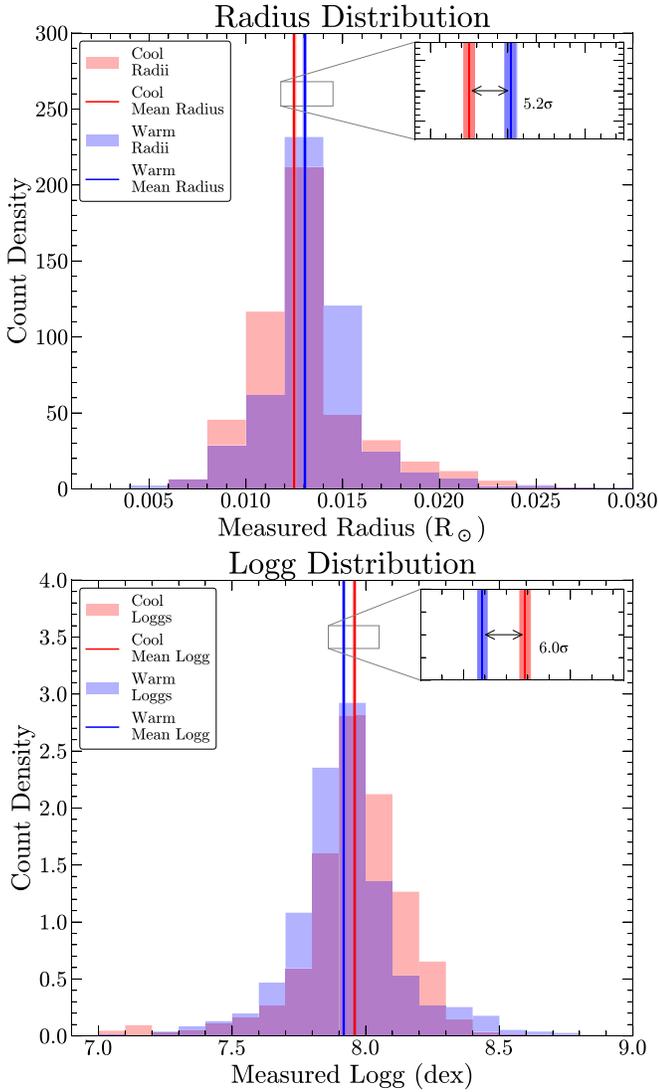

**Figure 6.** The distribution of measured radii (top) and surface gravities (bottom) for the cool (red) and warm (blue) subsamples of the combined catalog. The red and blue vertical lines correspond to the weighted means of the cool and warm distributions, respectively. The mean radius for the cool and warm samples is $0.01250 \pm 0.00007$ and $0.01304 \pm 0.00008\ R_\odot$, respectively. The mean surface gravity for the cool and warm samples is $7.958 \pm 0.005$ and $7.918 \pm 0.004$ dex, respectively.

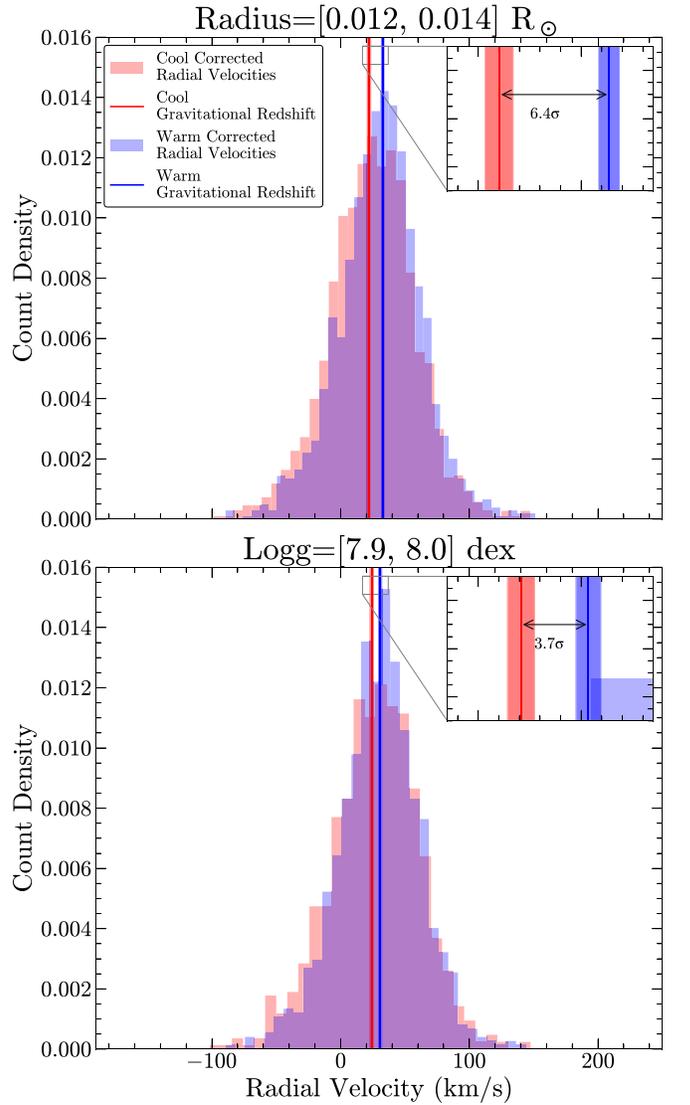

**Figure 7.** The difference in measured gravitational redshifts between the cool and warm subsamples of the combined catalog for redshifts measured by binning at the peak radius (top) and peak surface gravity (bottom) for one run of the bias correction simulation. Our measured gravitational redshifts for the cool and warm samples, including bias corrections and mean radius or surface gravity corrections, are shown in red and blue, respectively. The mean redshift for the cool and warm samples computed by binning in radius is $22.0 \pm 1.3$ and $32.7 \pm 1.0$ km s$^{-1}$, respectively. The mean redshift for the cool and warm samples computed by binning in surface gravity is $24.2 \pm 1.3$ and $30.7 \pm 1.2$ km s$^{-1}$, respectively. The LSR and asymmetric drift-corrected radial velocities in the peak bin for the cool and warm samples are shown in the red and blue histograms, respectively.

The peak of the surface gravity distribution occurs in the [7.9, 8.0] dex bin. There are 982 and 992 WDs in this bin in the cool and warm subsamples, respectively. In Figure 7, we show the LSR and asymmetric drift-corrected radial velocities in this peak surface gravity bin as well as the mean-surface gravity-corrected gravitational redshifts of each subsample. The mean surface gravity correction is performed in the same manner as the mean radius correction, using the La Plata models to adjust each gravitational redshift to the overall mean surface gravity of both samples for a more direct comparison. With this correction, the mean redshift for the cool and warm samples computed by binning in surface gravity is $24.2 \pm 1.3$ and $30.7 \pm 1.2 \,\mathrm{km\,s^{-1}}$, respectively, for this run of the bias correction simulation. This corresponds to a significance of the difference between the measured gravitational redshifts of $3.7\sigma$ in the correct direction predicted by theory. Before applying this correction, the significance of the difference for this bias-correction run is $3.6\sigma$ in the same direction, with cooler WDs having smaller gravitational redshifts at a fixed surface gravity. Across all bias simulation runs, the significance of the corrected gravitational redshifts varies between $3.6\sigma$ and $3.7\sigma$. For the uncorrected redshifts, this significance varies between $3.5\sigma$ and $3.6\sigma$. This is strong evidence for the temperature dependence of WD structure at the peak of the surface gravity distribution. The minimum significance of this detection is $3.6\sigma$, and the median significance of this detection is $3.7\sigma$ for the mean-surface gravity-corrected redshifts.

We emphasize that measurements of photometric radii are more robust against the underlying assumptions of model spectra than measurements of spectroscopic surface gravities. So, we consider the detection using measured radii to be the best indicator of the existence of temperature-dependent WD structure. This detection is supported by the results using spectroscopic surface gravities.

### 4.2. Comparison to Previous Studies

In this work, we have directly measured the temperature dependence of the WD mass–radius relation by empirically determining the relation at two different temperature ranges. This method expands upon other studies that have implicitly measured the temperature dependence of the mass–radius relation or have measured the effects of a combination of temperature and other factors. By considering WDs in visual binaries (H. E. Bond et al. 2015, 2017a, 2017b), in eclipsing binaries (S. G. Parsons et al. 2017), in wide binaries (S. Arseneau et al. 2024), or with astrometric microlensing data (K. C. Sahu et al. 2017; P. McGill et al. 2023), many studies have produced high-precision mass, radius, and temperature determinations of individual WDs. For these high-quality measurements, individual objects can be used to constrain the dependence of the WD mass–radius relation on both temperature and hydrogen envelope thickness. These parameters are degenerate when considering individual objects, as both warm WDs and WDs with thick hydrogen layers have a larger radius at a given mass compared to cool or thin hydrogen layer WDs (G. Fontaine et al. 2001). S. G. Parsons et al. (2017) ruled out thin hydrogen atmospheres by comparing their measurements to theoretical mass–radius relations. S. Arseneau et al. (2024) have a sufficiently large sample of binaries that they could bin their WD measurements in temperature, but they did not quite have the sample size to yield a statistically significant detection of the temperature dependence of the mass–radius relation.

In our work, we are unable to measure the gravitational redshifts as directly as some of these in-depth individual studies, where the spatial velocity measurement is available from the binary companion and therefore can be isolated from the gravitational redshift. But thanks to the large number of objects in our study, we can expand upon the methods of V. Chandra et al. (2020b) and average over random individual motions in the Galaxy and over the distribution of hydrogen layer thickness to empirically measure the WD mass–radius relation. V. Chandra et al. (2020b) used only 3316 WDs in their final sample, too small of a sample to detect the temperature dependence of the mass–radius relation. With our larger sample, we are able to directly detect the temperature dependence of the WD mass–radius relation.

A. Gianninas et al. (2010), A. Bédard et al. (2020), and other works have implicitly detected the dependence of the mass–radius relation on temperature by plotting observed WD surface gravities as a function of temperature. In these plots, the surface gravity distribution of hotter WDs is shifted toward lower surface gravities than the corresponding distribution for cooler stars, in line with expectations from theory. However, since no empirical mass–radius relation can be obtained from surface gravity measurements alone, these methods are not direct measurements of the effect of temperature on the WD mass–radius relation. Rather, they indicate that, on average, the hotter WDs have larger radii, under the assumption that the underlying mass distributions in the hotter and cooler subsamples are the same. Thus, our direct empirical determination of the temperature dependence of the WD mass–radius relation expands upon these previous works.

### 5. Discussion and Conclusions

In this paper, we have used catalogs of measured radial velocities, surface gravities, effective temperatures, and radii for all DA WDs observed in previous generations of SDSS and in the 19th Data Release of SDSS-V. A companion paper (N. R. Crumpler et al. 2024, in preparation) will describe the construction and validation of this catalog in detail. The companion paper, catalogs, and code used to produce both the companion paper and this work will be released when the 19th Data Release of SDSS-V is made publicly available. We use the measurements from these catalogs to isolate WD gravitational redshifts as a function of surface gravity or radius, providing constraints on the WD mass–radius relation that are largely independent of the assumptions underlying model spectra.

We compare our measured mass–radius relation to the La Plata models, which vary in core composition and hydrogen layer thickness in accordance with theory and observation, and find our results are consistent with theory. We divide our catalogs into cool ($<11{,}000\,\mathrm{K}$) and warm ($>16{,}000\,\mathrm{K}$) subsamples and remeasure the mass–radius relation. We find strong evidence that the mass–radius relation of WDs depends on their effective temperatures in our gravitational redshifts computed from both photometric radii and spectroscopic surface gravities. The location of the peak in the radius and surface gravity distributions varies depending on the temperature of the sample. The mean radius of the cool sample is smaller by $5.2\sigma$, and the mean surface gravity of the cool sample is larger by $6.0\sigma$ than that of the warm sample. The peak of either sample should occur at the same fixed mass. This variation is in line with expectations from theory, since a cool WD is expected to have a smaller radius and thus a larger surface gravity than a WD of the same mass at a warmer temperature.

We also investigate the difference in gravitational redshift within the peak bin of the radius and surface gravity distributions. At the peak of the WD radius distribution, the median significance of the temperature dependence in gravitational redshift variation is $6.4\sigma$ across all bias correction simulations. At the peak of the WD surface gravity distribution, the median significance of the temperature dependence in gravitational redshift variation is $3.7\sigma$. In both cases, the direction of the temperature dependence is in agreement with our expectations from theoretical models.

This analysis suffers from uncertainties tied to additional sources of scatter in our measurements. Some of these sources of scatter are physical, such as the dependence of the WD mass–radius relation on the WD core and atmospheric composition. The proportion of hydrogen to helium in the atmosphere and the thickness of the atmosphere are not well known and may vary among WDs (J. L. Provencal et al. 1998). The structure of the WD varies along with these parameters, adding a source of dispersion to the mass–radius relation. For example, for a 15,000 K WD with a 0.013 $R_\odot$ radius, thick hydrogen layer models ($M_H/M_{WD} = 10^{-4}$; G. Fontaine et al. 2001) predict a mass of 0.61 $M_\odot$, while thin hydrogen layer models ($M_H/M_{WD} = 10^{-10}$; G. Fontaine et al. 2001) predict a mass of 0.57 $M_\odot$. The most common WDs are carbon-oxygen WDs, and for these objects, the proportion of carbon to oxygen in the core is not fully known and may vary among stars (J. Isern et al. 2013). The effects of core composition are subdominant to the effects of the envelope composition (P. McGill et al. 2023).

Additionally, a large source of scatter in our measurements comes from the dependence of our parameters on the spectrum SNR. Most SDSS WD observations have SNRs < 10, too low to be useful in this analysis, reducing the statistical power of our sample. Even for WD spectra with SNR > 10, the uncertainties on the measured spectroscopic parameters are still fairly large. In an ideal sample, all coadded spectra would have SNR > 20 in order to be at the limiting precision of our measurement procedures. The high noise in our measurements due to poor SNR increases the dependence of our results on bias corrections far from the peak in the WD radius or surface gravity distributions. Decreasing this noise by increasing SNR will be vital in obtaining a confidently measured mass–radius relation over the whole of WD radius or surface gravity parameter space.

Future work can extend this analysis to better constrain the dependence of the WD mass–radius relation on effective temperature. SDSS-V is ongoing and is predicted to make tens of thousands more observations of WDs before the end of the survey. This increased sample size will continue to improve the available statistics characterizing the population properties of DA WDs. Our constraints on the temperature dependence of WD structure at both low and high WD masses are weak. In both of these regimes, the number density of WDs is much lower relative to WDs near the peak of the mass distribution, making it difficult to obtain sufficient observations to perform statistical analyses with high constraining power. Furthermore, sufficiently low-mass WDs must form via binary evolution and require more careful treatment than is done in this analysis to accurately measure the physical parameters of the star. Additional, targeted observations of both low- and high-mass WDs combined with analysis methods tailored to the high binary fraction of low-mass WDs would help to better characterize the temperature dependence across the full mass–radius or mass–surface gravity relation. Our understanding of the sharp-peak bias correction, especially when applied to gravitational redshift measurements from binning in surface gravity, could use a better characterization of deeper uncertainties. An alternative to the sharp-peak bias correction would be to perform this analysis using a method that does not require binning, such as kernel density estimation. As our measurements of the empirical WD mass–radius relation as a function of temperature continue to improve, these measurements should be able to test theoretical models and potentially discover new physical effects in the structure and evolution of WD stars.

## Acknowledgments

N.R.C. is supported by the National Science Foundation Graduate Research Fellowship Program under grant No. DGE2139757. Any opinions, findings, and conclusions or recommendations expressed in this material are those of the author and do not necessarily reflect the views of the National Science Foundation. V.C. gratefully acknowledges a Peirce Fellowship from Harvard University. N.L.Z. acknowledges support by the JHU President's Frontier Award and by the seed grant from the JHU Institute for Data Intensive Engineering and Science. S.A. was supported by the JHU Provost's Undergraduate Research Award. C.B. acknowledges support from NSF grant AST-2307865.

Funding for the Sloan Digital Sky Survey V has been provided by the Alfred P. Sloan Foundation, the Heising-Simons Foundation, the National Science Foundation, and the Participating Institutions. SDSS acknowledges support and resources from the Center for High-Performance Computing at the University of Utah. SDSS telescopes are located at Apache Point Observatory, funded by the Astrophysical Research Consortium and operated by New Mexico State University, and at Las Campanas Observatory, operated by the Carnegie Institution for Science. The SDSS website is www.sdss.org.

SDSS is managed by the Astrophysical Research Consortium for the Participating Institutions of the SDSS Collaboration, including the Carnegie Institution for Science, Chilean National Time Allocation Committee (CNTAC) ratified researchers, Caltech, the Gotham Participation Group, Harvard University, Heidelberg University, The Flatiron Institute, The Johns Hopkins University, L'Ecole polytechnique fédérale de Lausanne (EPFL), Leibniz-Institut für Astrophysik Potsdam (AIP), Max-Planck-Institut für Astronomie (MPIA Heidelberg), Max-Planck-Institut für Extraterrestrische Physik (MPE), Nanjing University, National Astronomical Observatories of China (NAOC), New Mexico State University, The Ohio State University, Pennsylvania State University, Smithsonian Astrophysical Observatory, Space

Telescope Science Institute (STScI), the Stellar Astrophysics Participation Group, Universidad Nacional Autónoma de México, University of Arizona, University of Colorado Boulder, University of Illinois at Urbana-Champaign, University of Toronto, University of Utah, University of Virginia, Yale University, and Yunnan University.

This work has made use of data from the European Space Agency (ESA) mission Gaia (https://www.cosmos.esa.int/gaia), processed by the Gaia Data Processing and Analysis Consortium (DPAC, https://www.cosmos.esa.int/web/gaia/dpac/consortium). Funding for the DPAC has been provided by national institutions, in particular the institutions participating in the Gaia Multilateral Agreement.

*Software:* astropy (Astropy Collaboration et al. 2013, 2018, 2022).

### ORCID iDs

Nicole R. Crumpler ● https://orcid.org/0000-0002-8866-4797
Vedant Chandra ● https://orcid.org/0000-0002-0572-8012
Nadia L. Zakamska ● https://orcid.org/0000-0001-6100-6869
Gautham Adamane Pallathadka ● https://orcid.org/0000-0002-5864-1332
Stefan Arseneau ● https://orcid.org/0000-0002-6270-8624
Nicola Gentile Fusillo ● https://orcid.org/0000-0002-6428-4378
J. J. Hermes ● https://orcid.org/0000-0001-5941-2286
Carles Badenes ● https://orcid.org/0000-0003-3494-343X
Priyanka Chakraborty ● https://orcid.org/0000-0002-4469-2518
Boris T. Gänsicke ● https://orcid.org/0000-0002-2761-3005
Stephen P. Schmidt ● https://orcid.org/0000-0001-8510-7365